\documentstyle[prl,preprint,aps]{revtex}
\begin{document}
\draft
\title{Effects of isospin-symmetry violation on tests of the standard
model using parity-violating electron scattering}
\author{W.~E. Ormand}
\address{Department of Physics and Astronomy, 202 Nicholson Hall,\\
Louisiana State University, Baton Rouge, LA 70803-4001}
\maketitle
\begin{abstract}
The effects of isospin-symmetry breaking on the observables 
for parity-violating electron scattering are investigated within the
framework of the nuclear shell model for  
$^{12}$C, $^{16}$O, and $^{28}$Si. Contributions due
to mixing with low-lying states as well as admixtures of $1p-1h$ 
configurations (via the radial wave functions) are 
accounted for. It is found that isospin-mixing  
can be important, and the consequences regarding
precision tests of the standard model are addressed.
\end{abstract}
\pacs{PACS number(s): 24.80.+y, 25.30.Bf }


One of the most successful theories in 
physics is the standard model for the electroweak 
interaction~\cite{r:standard}. Because
of its extraordinary success and the fact that the 
origin of many of the parameters defining it are not that well understood,
the focus of many programs is to perform high-precision experiments 
with the hope of discovering fingerprints of physics beyond the standard
model. Atomic nuclei provide a convenient laboratory for these tests, 
although in some cases nuclear-structure effects must be
accounted for. Two examples are:
(1) the $ft$ values for superallowed Fermi beta
decay~\cite{r:beta1,r:beta2,r:beta2a,r:beta3}, 
which test the conserved vector 
current hypothesis~\cite{r:cvc} and the unitarity of the 
Cabibbo-Kobayashi-Maskawa matrix~\cite{r:ckm}; 
and (2) parity violation in electron 
scattering from even-even, $N=Z$ nuclei, which offers a window into the 
neutral-current sector of the weak 
interaction~\cite{r:pv1,r:Musolf,r:musolf_a}. 
In both examples is that the systems were chosen to minimize the 
effects due to the internal structure of the nucleus. 
In fact, if isospin is a good quantum number, the measured 
observables are hardly affected by nuclear structure at all. 
However, because of 
the presence of the Coulomb interaction and 
charge-dependent components in the strong interaction, 
isospin symmetry is violated and corrections can be 
expected. In the case of Fermi beta decay, the corrections are small 
($\le 0.7$\%), but  
important~\cite{r:beta1,r:beta2,r:beta2a,r:beta3}. 
In the case of parity-violating electron scattering, the goal is 
to perform a measurement that provides a 1\% test of the standard model. 
Effects due to isospin-symmetry breaking on the parity-violation 
observables have been estimated previously~\cite{r:donnelly}, 
with the conclusion being that for $^{12}$C  
isospin-mixing corrections are expected to be less than 
1\% for essentially all momenta transfer.

With a series of experiments now planned for TJLAB, 
this issue is revisited using improved nuclear models. 
The principal improvements implemented are: 
(1) using radial wave functions obtained from Hartree-Fock (or Woods-Saxon) 
calculations with separation energies determined from intermediate states of 
the $A-1$ system to account for the mixing between the ground state and 
one particle-one hole ($1p-1h$) excitations and 
(2) performing the shell-model 
calculations in proton-neutron formalism while including empirically 
determined isospin-nonconserving (INC) interactions~\cite{r:ormand}. The 
advantages in (2) are that the interaction inducing 
isospin mixing is constrained to reproduce binding energy differences 
within isospin multiplets and that the 
sum over states that can mix into the ground state is carried out 
implicitly. For (1), an important action of the Coulomb
force is to introduce admixtures of $1p-1h$ states that effectively 
renormalize the proton radial wave functions relative to the 
neutrons by decreasing the single-particle separation energy. 
Although these effects are in principal included in the definition of the 
Hartree-Fock mean field, it is important to correctly account for 
the separation energy of the single-particle states relative to the 
complete set of states in the $A-1$ system. 
These procedures are in contrast to those of 
Ref.~\cite{r:donnelly}, where a more qualitative result was obtained 
primarily through a two-level model and adopting  
a ``worst-case'' philosophy assuming isospin-mixing matrix elements
of the order 300~keV. 

The observable of interest for elastic scattering from an
even-even, $N=Z$ target is the 
parity-violating electron scattering asymmetry~\cite{r:donnelly,r:asymm}
\begin{equation}
{\cal A}=
\frac{{\rm d}\sigma^+-{\rm d}\sigma^-}{{\rm d}\sigma^++{\rm d}\sigma^-} = 
-\left(\frac{G_Fq^2}{4\pi\alpha\sqrt{2}}\right)
\frac{\tilde F_{C}(q)}{F_{C}(q)},
\label{e:assymetry}
\end{equation}
where $\pm$ refers to electrons with helicity $\pm 1$, $G_F$ is the weak 
interaction Fermi constant, $\alpha$ is the fine structure constant, and
$q=|{\bf q}|$ is the magnitude of the three momentum transfer.
The dependence on nuclear structure is embodied in the Coulomb 
form factors $F_C(q)$ and $\tilde F_C(q)$ for the electromagnetic and
neutral currents, respectively. Both form factors
have the same form, with the primary difference only being in the charges.
In particular, the electromagnetic form factor is 
given by 
\begin{equation}
F_C(q)=\sum_{\mu}^{protons} n_\mu^pf_\mu^p(q)+
\sum_{\mu}^{neutrons} n_\mu^nf_\mu^n(q),
\label{e:form}
\end{equation}
where $\mu$ denotes the labels for each single-particle orbit,
$n_\mu^{p(n)}$ the number of protons(neutrons) occupying
each single-particle orbit, and $f_\mu^{p(n)}(q)$ is given by 
\begin{equation}
f_\mu^{p(n)}(q)=
\int_0^\infty r^2{\rm d}r(R_\mu^{p(n)}(r))^2\hat M_0^{p(n)}(qr), 
\label{e:sp_form}
\end{equation}
where $R_\mu^{p(n)}(r)$ is the radial wave function, and 
$\hat M_0^{p(n)}(qr)$ is a generalized 
charge operator given by
\begin{equation}
\hat M_0^{p(n)}(qr) = \frac{1}{\sqrt{4\pi}}\left[ 
\frac{G_E^{p(n)}(\tau)}{\sqrt{1+\tau}}j_0(qr)
+(G_E^{p(n)}(\tau)-2G_M^{p(n)}(\tau))
\frac{2\tau j_1(qr)}{qr}{\bf \sigma}\cdot {\bf l}\right]
\label{e:charge}
\end{equation}
where $\tau=q^2/4m_N^2$, $m_N$ is the mass of the nucleon, and 
$G_E^{p(n)}$ and $G_M^{p(n)}$ are the Sachs~\cite{r:Sachs}
electric and magnetic intrinsic form factors. 
Here, the dipole forms of Ref.~\cite{r:Galster} are used.
For the neutral current, $\tilde F_C(q)$ is also
given by Eqs.~(\ref{e:form})-(\ref{e:charge}) 
using the intrinsic weak form factors~\cite{r:Musolf}
\begin{eqnarray}
G_{E(M),W}^p&=&
[(1-4\sin^2\theta_W)G_{E(M)}^p-G_{E(M)}^n ] \nonumber \\
G_{E(M),W}^n&=&
[-G_{E(M)}^p+(1-4\sin^2\theta_W)G_{E(M)}^n ], 
\end{eqnarray}
with $\sin^2\theta_W\sim 0.230\pm 0.005$\cite{r:theta_W}. 
Lastly, a correction due to the center-of-mass
must also be applied. For a
harmonic oscillator potential, this is well 
defined~\cite{r:Tassie}, and Eq.~(\ref{e:form}) is multiplied by
$\exp(a_0^2q^2/4A)$, where $a_0^2=m_N\omega/\hbar$ is the oscillator
parameter, which may be accurately parameterized with 
$\hbar\omega=45A^{-1/3}-25A^{-2/3}$~MeV. Although this simple 
decomposition is not possible for the more realistic potentials 
used in Woods-Saxon or Hartree-Fock calculations, the 
harmonic-oscillator result represents a reasonable approximation. 

In the limit that isospin is a good quantum number, $n_\mu^p=n_\mu^n$
and $R_\mu^p(r)=R_\mu^n(r)$, and Eq.~(\ref{e:assymetry}) reduces to
\begin{equation}
{\cal A}_0=
[G_Fq^2/\pi\alpha\sqrt{2}]\sin^2\theta_W=
3.22\times10^{-6}q^2,
\label{e:reduce}
\end{equation}
with $q^2$ measured in fm$^{-2}$. 
It is this simple form that makes
experiments on even-even, $N=Z$ nuclei an attractive choice for 
testing the standard model, as any deviation from the simple
$q^2$ dependence might be a signature of physics beyond the standard model.
On the other hand, isospin is not 
a conserved quantity and corrections to Eq.~(\ref{e:reduce}) must be
accounted for. These corrections may be embodied in 
the factor $\Gamma(q)$~\cite{r:donnelly} defined as
\begin{equation}
{\cal A}={\cal A}_0(1+\Gamma(q)),
\end{equation}
which, from Eq.~(\ref{e:assymetry}), may be written as
\begin{equation}
\Gamma(q)=-[1+\tilde F_C(q)/2\sin^2\theta_W F_C(q)].
\label{e:correction}
\end{equation}

An estimation of $\Gamma(q)$ begins by noting that as in the case of 
superallowed Fermi beta decay, two types of isospin mixing must be accounted
for~\cite{r:beta2}. The first contribution is due to mixing 
between states contained within the shell-model configuration space.
For example, the $^{12}$C model space consists of the
$0p_{3/2}$ and $0p_{1/2}$ orbitals ({\it p}-shell), 
and there are 5, 2, and 2 configurations
leading to $J^\pi=0^+$ and $T=0$, 1, and 2, respectively. 
The INC interaction is composed of 
isospin operators of rank zero, one, and two, and 
consequently mixes together all the $J^\pi=0^+$ 
states in $^{12}$C. In regards to Eq.~(\ref{e:form}),  
isospin mixing within the configuration space  
leads to $n_\mu^p\ne n_\mu^n$.

In addition to the mixing between states within
the configuration space, mixing with states that lie outside
the model space must also be taken into account. The Coulomb
interaction can strongly mix $1p-1h$, $2\hbar\Omega$
excitations, eg., $0p_{3/2}\rightarrow 1p_{3/2}$, into the ground state. 
For Fermi transitions, these excitations were 
accounted for by examining differences in the proton and neutron 
radial wave functions~\cite{r:beta2,r:beta2a,r:beta3}. 
For closed-shell 
configurations, the mixing between the ground state and 
$1p-1h$ states is contained within the   
Hartree-Fock (HF) procedure, and an estimate of
this contribution might be obtained by evaluating the 
$f_\mu^{p(n)}$ using HF
radial wave functions. 

As a start, $^{12}$C is examined in detail. The occupation 
factors $n_\mu^{p(n)}$ were obtained from shell-model calculations 
carried out in proton-neutron formalism within the {\it p}-shell. 
In addition to the isospin-conserving shell-model Hamiltonian   
CKPOT~\cite{r:ckpot}, isospin mixing within the configuration space was
accounted for by including the INC
Hamiltonian of Ref.~\cite{r:ormand}. The INC interaction contains both 
Coulomb and charge-dependent nucleon-nucleon terms whose strengths were 
determined by the requirement that binding energy differences within isospin
multiplets be reproduced.
Interactions of this form have subsequently been used to 
examine isospin-mixing 
corrections to Fermi beta decay~\cite{r:beta3}, 
isospin-forbidden particle emission~\cite{r:part}, 
isospin-forbidden Fermi beta decay~\cite{r:forbidden}, 
and to predict the location of the proton-drip line~\cite{r:drip}.
For the $f_\mu(q)$, radial wave functions obtained from 
HF calculations utilizing Skyrme-type 
interactions~\cite{r:skyrme} and the shell-model occupation factors to 
define the HF densities were used.
Here, the M$^*$~\cite{r:skyrme_m} interaction was used, 
but others produced qualitatively the same results. 
In Fig.~\ref{fig:fig1}a, the square
of the charge form factor obtained with HF radial wave functions
(dotted line) is compared with experimental data~\cite{r:c12data}.
The HF result reproduces the experimental data up
to the diffraction minimum, but is in serious disagreement beyond. 
This region
is quite sensitive to the details of the radial wave 
functions, and in particular the separation energy, which can be 
obtained from the experimental binding energies 
for $^{11}$B and $^{11}$Cs~\cite{r:audi}.
Towards this end, the same procedure used to evaluate the 
radial-overlap correction for Fermi beta 
decay~\cite{r:beta2,r:beta2a} may be used. 
The occupation factors may be rewritten as
\begin{equation}
n_\mu=\sum_\pi{\textstyle {1\over2}}S(\mu,\pi),
\end{equation} 
where the sum extends over the complete set of states $|\Psi(\pi)\rangle$ of
the intermediate $A-1$ system, and the spectroscopic factor $S(\mu,\pi)$ is
given by
\begin{equation}
S(\mu,\pi)=\frac{|\langle \Psi(^{12}C)|a^\dagger_\mu |\Psi(\pi)\rangle|^2}
{2J_\pi+1}.
\end{equation}
The $f_\mu$ are then evaluated for each intermediate state by scaling the 
central part of the mean-field potential to yield the correct  
separation energy between the $^{12}$C
ground state and the intermediate state $|\Psi(\pi)\rangle$~\cite{r:audi}. 
The corresponding
charge form factor is illustrated by the solid line in Fig.~\ref{fig:fig1}a,
where it seen that much better agreement with experiment is achieved.

Shown in Fig.~\ref{fig:fig1}b
is the expected parity-violating asymmetry (solid line) compared with the
``pure'' standard model expectation (dashed line). In Fig.~\ref{fig:fig1}c, 
$|\Gamma(q)|$ is plotted (solid line), and is seen to increase 
rapidly; exceeding the critical value of 1\% at 
approximately 0.9~fm$^{-1}$. 
The large increase in 
$\Gamma(q)$ is due to the Coulomb 
interaction ``pushing'' the proton distribution out relative to the 
neutrons, leading to a shift in the diffraction minima between the 
Coulomb and weak form factors. At the diffraction minimum, 
$\Gamma(q)$ changes
sign, and actually crosses through zero again in a relatively flat region 
near the second maximum. This is a general feature that might be 
exploited for experiments on heavier nuclei. 

Also shown in Fig.~\ref{fig:fig1}c, is the relative dominance of 
the contribution to $\Gamma(q)$ due to the radial wave function.
The dashed line 
represents $\Gamma(q)$ evaluated using isospin-conserved occupation factors 
in conjunction with proton and neutron
radial wave functions, while the dotted
line shows the contribution due to isospin mixing within the shell-model
space (begins negative), and was evaluated 
by replacing $f_\mu^p(q)$ with $f_\mu^n(q)$.
 
$\Gamma(q)$ was also evaluated for  
$^{16}$O and $^{28}$Si. For $^{16}$O, the closed {\it p}-shell was assumed,
while for $^{28}$Si, shell-model
calculations were carried out in proton-neutron formalism within the 
{\it sd}-shell ($0d_{5/2}$, $0d_{3/2}$, and $1s_{1/2}$ orbitals) model space 
using the isospin-conserving USD Hamiltonian of 
Wildenthal~\cite{r:usd} and the 
{\it sd}-shell INC interaction in Ref.~\cite{r:ormand}. 
For $^{16}$O and $^{28}$Si, the effects of the intermediate 
$A-1$ states were taken into account. 
The resulting
values of $|\Gamma(q)|$ are shown in Fig.~\ref{fig:fig2} and compared with 
$^{12}$C. The 1\% value is illustrated by the 
dot-dashed line.

Another nucleus of interest is $^4$He. 
Because of the small charge, 
isospin-mixing corrections are expected to be small. 
For completeness, $\Gamma(q)$ was also evaluated for $^4$He assuming a 
closed $0s_{1/2}$ core, and is 
illustrated in Fig.~\ref{fig:fig2}. Although
$\Gamma(q)$ is found to be small, and in overall agreement with
the results of Ref.~\cite{r:he4_isospin}, a 
weakness in the calculation is the position of the diffraction 
minimum, which illustrates the need to go beyond the closed $0s_{1/2}$ 
configuration and to include meson-exchange currents~\cite{r:he4_mus}.
Another approach currently under investigation, is to 
perform a large-basis, no-core shell-model calculation utilizing a 
realistic interaction~\cite{r:nav96} that also 
includes INC components, as was recently done for the superallowed Fermi 
transition in $^{10}$C~\cite{r:navratil}. In this case, 
both contributions to isospin mixing are contained within the same formalism.  
This calculation would be an excellent compliment 
to earlier works that utilized variational $^4$He ground-state wave 
functions~\cite{r:he4_mus,r:he4_isospin}.

In addition to providing a test of the standard model,   
parity-violating electron scattering  
is also sensitive to any strangeness content in the 
nucleon~\cite{r:strange}, and can be used as a probe to measure the
strangeness form factor. These experiments will be carried out at
higher momenta transfer, and from 
Fig.~\ref{fig:fig2} it is clear that they
have to be carefully designed in order to minimize  
effects due to nuclear structure. Towards this end, it is likely that  
the cross over through zero exhibited in $\Gamma(q)$ near the 
second maximum of $F_C^2(q)$ may be exploited.

In conclusion, effects due to isospin-mixing on the observables for 
parity-violating electron scattering were evaluated for $^{12}$C,
$^{16}$O, and $^{28}$Si. The method employed very 
closely mirrors that used for superallowed Fermi beta decay. In particular, 
the influence of the intermediate parent states in the $A-1$ system was found
to be important for a proper description of the charge form factor. 
In general,
it is found that $\Gamma(q)$ increases rapidly with $q$, and often
exceeds the critical value of 1\%.  
However, $\Gamma(q)$ is also found to ``cross-over'' and pass through zero
in a region just past the second maximum in the charge form factor. It might 
be possible to exploit this feature to make experiments on 
heavier nuclei amenable to tests of the weak interaction or as a 
probe of the strangeness content in the nucleon.

Discussions with M.~J. Ramsey-Musolf and W.~Haxton 
are gratefully acknowledged. This works stems from the
fifth annual JLAB/INT workshop ``Future directions in parity violation''
held at the Institute for Nuclear Theory.
Support from NSF Cooperative agreement No. EPS~9550481, 
NSF Grant No. 9603006, and DOE contract
DE--FG02--96ER40985 is acknowledged.

\bibliographystyle{try}

\begin{thebibliography}{11} 

\bibitem{r:standard} S. Weinberg, Phys. Rev. {\bf 19}, 1264 (1967);
A.~Salam, in {\sl Elementary particle physics}, ed. N.~Svartholm
(Stokholm, 1968), p. 367; S.~L.~Glashow, J.~Iliopoulos, and L.~Maiani,
Phys. Rev. {\bf D} 2, 1285 (1970).

\bibitem{r:beta1} J.~C.~Hardy {\it et al.}, 
Nucl. Phys. {\bf A509}, 429 (1990); 
G.~Savard {\it et al.}, Phys. Rev. Lett. {\bf 42}, 1521 (1995).

\bibitem{r:beta2} I.~S.~Towner, J.~C.~Hardy, and M.~Harvey, Nucl. Phys. 
{\bf A284}, 269 (1977).

\bibitem{r:beta2a} W.~E.~Ormand and B.~A.~Brown, Nucl. Phys. {\bf A440}, 
274 (1985).

\bibitem{r:beta3} W.~E.~Ormand and B.~A.~Brown, Phys. Rev. 
Lett. {\bf 62}, 866 (1989)

\bibitem{r:cvc} S.~S. Gerstein and Y.~B.~Zeldovich, Sov. Phys. Jetp.
{\bf 2}, 576 (1956); R.~P.~Feynman and M.~Gell~Mann, Phys. Rev. 
{\bf 109}, 193 (1958).

\bibitem{r:ckm}  N.~Cabibbo, Phys. Rev. Lett. {\bf 10}, 531 (1963);
M.~Ko\-ba\-ya\-shi and T.~Maskawa, Prog. Theor. Phys. {\bf 49}, 652 (1973). 

\bibitem{r:pv1} {\sl Parity violation in
electron scattering}, ed. E.~J.~Beise and R.~D.~Mckeown 
(World Scientific, Singapore, 1990)

\bibitem{r:Musolf} M.~J.~Musolf and T.~W.~Donnelly, Nucl. Phys. 
{\bf A546}, 509 (1992).

\bibitem{r:musolf_a} M.~J.~Musolf {\it et al.}, Phys. Rep. {\bf 239}, 
1 (1994).

\bibitem{r:donnelly} T.~W. Donnelly, J.~Dubach, I.~Sick, Nucl. Phys. 
{\bf A503}, 589 (1989); T.~W. Donnelly, in Ref.~\cite{r:pv1}, p. 192.

\bibitem{r:ormand} W.~E. Ormand and B.~A.~Brown, Nucl. Phys. 
{\bf A491}, 1 (1989).

\bibitem{r:asymm} G.~Feinberg, Phys. Rev. {\bf D} 12, 3575 (1975);
J.~D. Walecka, Nucl. Phys. {\bf A285}, 349 91977); T.~W.~Donnelly and
R.~D.~Peccei, Phys. Reports {\bf 50}, 1 (1979).

\bibitem{r:Sachs} R.~G.~Sachs, Phys. Rev. {\bf 126}, 2256 (19962).

\bibitem{r:Galster} S.~Galster {\it et al.}, Nucl. Phys. {\bf B32},
221 (1971).

\bibitem{r:theta_W} U. Amaldi {\it et al.}, Phys. Rev. {\bf D} 36, 
1385 (1985).

\bibitem{r:Tassie} L.~J.~Tassie and F.~C.~Barker, Phys. Rev. {125}, 1034
(1958).

\bibitem{r:ckpot} S.~Cohen and D.~Kurath, Nucl. Phys. {\bf 73}, 1 (1965).

\bibitem{r:part} W.~E.~Ormand and B.~A.~Brown, Phys. Lett. B {\bf 174},
128 (1986).

\bibitem{r:forbidden} W.~E.~Ormand, Ph. D. thesis, 
(Michigan State University, East Lansing, 1986); 
 W.~E.~Ormand and B.~A. Brown, Phys. Rev. {\bf C} 52, 2455 (1995).

\bibitem{r:drip} W.~E.~Ormand, Phys. Rev. {\bf C} 53, 214 (1996);
W.~E.~Ormand, Phys. Rev. {\bf C} 55, 2407 (1997).

\bibitem{r:skyrme} T.~H.~R.~Skyrme, Phil. Mag. {\bf 1}, 1043 (1956);
D. Vautherin and D.~M. Brink, Phys. Rev. {\bf C} 5, 626 (1972);
C.~B. Dover and N.~V. Giai, Nucl. Phys. {\bf A190}, 373 (1972).

\bibitem{r:skyrme_m} J.~Bartel {\it et al.}, Nucl. Phys. {\bf A386}, 79 
(1982).

\bibitem{r:audi} G. Audi and A.~H.~Wapstra, Nucl. Phys. {\bf A565}, 1
(1993); {\it Table of Isotopes}, edited by R.~B.~Firestone and
V.~S.~Shirley, eighth edition (Wiley \& Sons, New York, 1996); In the absence 
of experimental data, theoretical excitation energies were used.

\bibitem{r:c12data} H. Uberall, {\sl Electron scattering from complex nuclei,
part A}, (Academic Press, New York, 1971). From Figure 3.20.

\bibitem{r:usd} B.~H.~Wildenthal, Prog. Part. Nucl. Phys., vol. 11, ed. 
D.~H.~Wilkinson (Pergamon, Oxford, 1984) p. 5.

\bibitem{r:he4_isospin} S.~Ramavataram, E.~Hadjimichael, and T.~W.~Donnely,
Phys. Rev. C{\bf 50}, 1175 (1994).

\bibitem{r:he4_mus} M.~J. Musolf, R.~Sciavilla, and T.~W.~Donnely,
Phys. Rev.~C{\bf 50}, 2173 (1994). 

\bibitem{r:nav96} P.~Navratil and B.~R.~Barrett, Phys. Rev. C{\bf54}, 2986 
(1996).

\bibitem{r:navratil} P.~Navratil, B.~R.~Barrett, and W.~E.~Ormand, Phys.
Rev. C{\bf 56}, 2542 (1997).

\bibitem{r:strange} D.~H.~Beck, Phys. Rev. {\bf D} 39, 3248 (1989).

\end{thebibliography}

\newpage

\begin{figure}
\caption{$^{12}$C results.
In (a), $F_C^2(q)$ is plotted 
and compared with experimental data. The dotted line represents the 
results obtained with the Hartree-Fock radial wave functions, 
while the solid line respresents the 
results obtained by summing over the intermediate $A-1$ states as 
explained in the text. 
In (b) and (c) the parity-violation 
assymmetry parameter, 
${\cal A}$, and isospin-mixing correction $|\Gamma(q)|$ (in \%, 
solid line), respectively, are plotted.
In (c), the contributions to $\Gamma(q)$ due to the differences in the 
radial wave functions and isospin mixing within the shell-model space 
are illustrated by the dashed and dotted lines, respectively.}
\label{fig:fig1}
\end{figure}
\begin{figure}
\caption{Calculated values of $|\Gamma(q)|$ (in \%) obtained for 
$^4$He (long dashed), $^{12}$C (solid), $^{16}$O (dotted), and 
$^{28}$Si (dashed).}
\label{fig:fig2}
\end{figure}

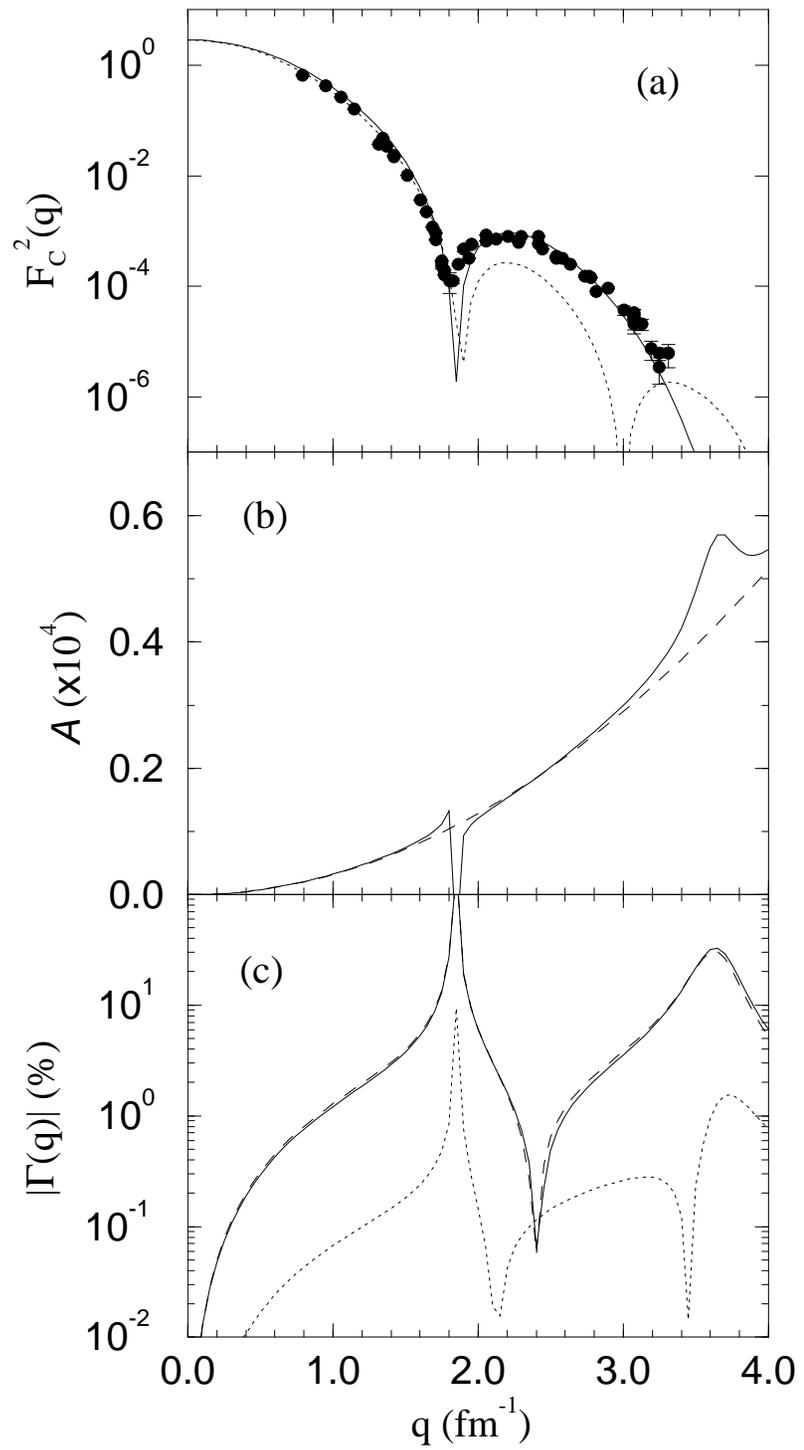
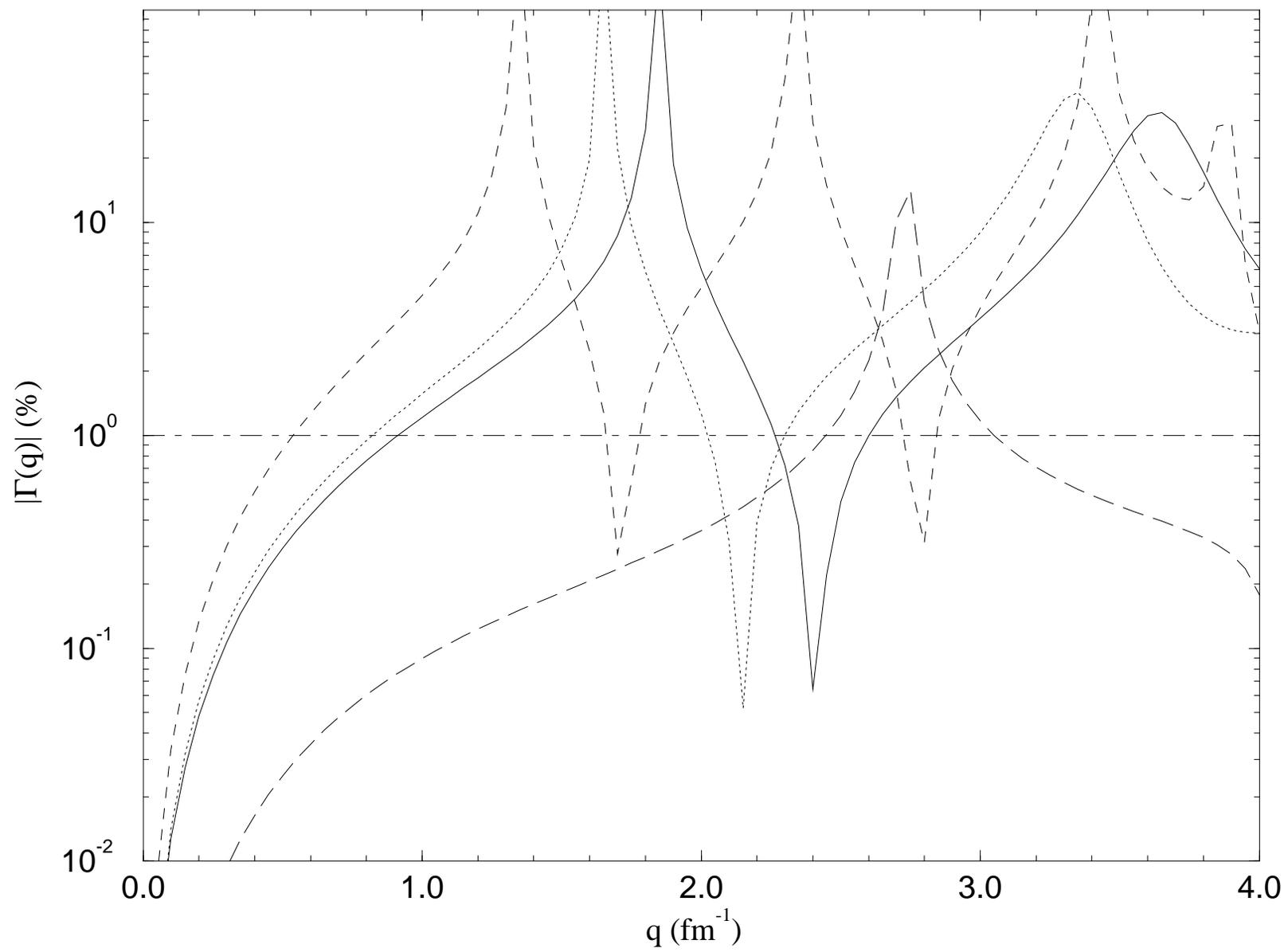
\end{document}